\documentclass{article} % For LaTeX2e
\usepackage[preprint]{neurips_2025}
\usepackage{bm}

\usepackage{hyperref}

\usepackage{algorithm}
\usepackage{algorithmic}
\usepackage{forloop}

\usepackage{multirow}
\usepackage{graphicx}
\usepackage{subfigure}
\usepackage{subcaption}
\usepackage{placeins}
\usepackage[utf8]{inputenc} % allow utf-8 input
\usepackage{fancyhdr}       % headers and footers
\usepackage[T1]{fontenc}    % use 8-bit T1 fonts
\usepackage{url}            % simple URL typesetting
\usepackage{booktabs}       % professional-quality tables
\usepackage{amsfonts}       % blackboard math symbols
\usepackage{nicefrac}       % compact symbols for 1/2, etc.
\usepackage{microtype}      % microtypography

\usepackage{amsmath}
\usepackage{amssymb}
\usepackage{mathtools}
\usepackage{amsthm}
\usepackage{tikz}

\usepackage[capitalize,noabbrev]{cleveref}

\newcommand{\newcommenter}[3]{%
    \expandafter\newcommand\csname #1\endcsname[1]{%
        \textcolor{#3}{[\textbf{#2}: ##1]}%
    }
}
\newcommenter{rahul}{Rahul}{green} % Example: Charlie's comments in green
\newcommenter{arka}{Arka}{red}

\theoremstyle{plain}
\newtheorem{theorem}{Theorem}
\newtheorem{theoremrestated}{Theorem} % Unnumbered environment for restatements (note the *)

\theoremstyle{definition}
\newtheorem{definition}{Definition}

\theoremstyle{remark}

\usepackage{xcolor}
\definecolor{mydarkblue}{rgb}{0,0.08,0.45}
\hypersetup{
colorlinks=true,
linkcolor=mydarkblue,
citecolor=mydarkblue,
filecolor=mydarkblue,
urlcolor=mydarkblue,
}

\title{An Attack to Break Permutation-Based Private Third-Party Inference Schemes for LLMs}

\author{%
Rahul Thomas$^{1,2}$ \quad
Louai Zahran$^{1}$ \quad
Erica Choi$^{1,3}$ \quad
Akilesh Potti$^{1}$ \quad
\AND
Micah Goldblum$^{1,3}$ \quad
Arka Pal$^{1}$\thanks{Project lead and corresponding author.} \\[1.5ex]
$^1$Ritual \quad $^2$Stanford University \quad $^3$Columbia University \\[0.5ex]
\texttt{\{rahulthomas, louai, erica, akilesh, micah, arka\}@ritual.net}
}

\begin{document}

\maketitle

\begin{abstract}

Recent advances in Large Language Models (LLMs) have led to the widespread adoption of third-party inference services, raising critical privacy concerns. Existing methods of performing private third-party inference, such as Secure Multiparty Computation (SMPC), often rely on cryptographic methods. However, these methods are thousands of times slower than standard unencrypted inference, and fail to scale to large modern LLMs. Therefore, recent lines of work have explored the replacement of expensive encrypted nonlinear computations in SMPC with statistical obfuscation methods - in particular, revealing permuted hidden states to the third parties, with accompanying strong claims of the difficulty of reversal into the unpermuted states. In this work, we begin by introducing a novel reconstruction technique that can recover original prompts from hidden states with nearly perfect accuracy across multiple state-of-the-art LLMs. We then show that extensions of our attack are nearly perfectly effective in reversing permuted hidden states of LLMs, demonstrating the insecurity of three recently proposed privacy schemes. We further dissect the shortcomings of prior theoretical `proofs' of permuation security which allow our attack to succeed. Our findings highlight the importance of rigorous security analysis in privacy-preserving LLM inference.
\end{abstract}

\section{Introduction}
\label{sec:introduction}

Recent advances in Large Language Model (LLM) capabilities have led to their widespread use for a diverse range of tasks \citep{zhu2024multilingualmachinetranslationlarge, kasneci2023chatgptgoodeducation, thirunavukarasu2023llms_medicine}. These models have demonstrated remarkable performance across domains including natural language processing, code generation, and complex reasoning tasks. However, modern LLMs are often very large -- sometimes comprising hundreds of billions of parameters -- necessitating significant hardware resources to deploy them for inference. Individuals and organizations therefore increasingly rely on third-party LLM inference services. This raises significant privacy implications, particularly in domains where confidentiality of data is paramount, such as healthcare, finance and legal applications, and in jurisdictions where data privacy is subject to regulations (e.g. GDPR in Europe). As such, a growing area of research interest is the creation of inference methodologies and schemes that protect the privacy of user prompts.

%In particular, recent \emph{open-weights} models demonstrate cutting-edge performance \citep{deepseekai2025deepseekr1incentivizingreasoningcapability, qwen2025qwen25technicalreport}, but remain difficult for many to run.

One approach to privacy-preserving-inference is based on having multiple parties participate jointly in performing the inference, with the idea that each party cannot itself reconstruct the input solely with the information that it is given in the protocol. This approach is known as Secure Multi-Party Computation (SMPC) and has a long history of application to general functions \citep{yao1982protocolssecurecomputations, goldreich1987howtoplay}. Recently, the methodologies of SMPC have been applied to LLMs \citep{huang2022cheetah, hao2022iron, pang2023bolt, akimoto2023privformer, dong2023pumasecureinferencellama7b, li2024nimbussecureefficienttwoparty}. A difficulty uniformly faced by these protocols is the computation of the many non-linearities present in transformer-based LLMs, which are not efficiently computable by standard SMPC approaches; most of the works attempt to ameliorate this by using piecewise polynomial approximations which are more well-suited for MPC algorithms. However, such approximation leads to degraded inference results, and remains more expensive than direct computation of the non-linearities.
%However, SMPC methods introduce significant computational and communication overhead, particularly so at non-linearities in the model.

Therefore, other works seek to mitigate the punitive costs of standard SMPC approaches by additionally utilizing statistical obfuscation approaches. In particular, recent work \citep{zheng2024permllmprivateinferencelarge, yuan2024securetransformerinferenceprotocol, luo2024centaurbridgingimpossibletrinity} has leveraged the permutation-equivariance properties of transformers \citep{xu2024permutationequivariancetransformersapplications} to propose permutation-based schemes for private inference. Under these schemes, hidden states are revealed as permuted plaintext to the party performing the inference. These works justify security by referring to the extremely large set of possibilities in the permutation space, and concluding that the reversal of these permuted states to the original user prompts is practically infeasible.

In this paper, we introduce a novel attack that is capable of reversing all permutation types used in the schemes above nearly perfectly into the original input tokens. First, we lay out its basic concept, and demonstrate its efficacy, in the unpermuted setting. We then extend the attack to the permuted setting. We discuss the key assumptions required for our attack, and discover that LLM hidden states strongly satisfy the property of various forms  of \emph{non-collision} which enable the high success rate of our attack.  We further dissect why the theoretical results of \citet{zheng2022towards,yuan2024securetransformerinferenceprotocol, luo2024centaurbridgingimpossibletrinity} on the security of permutations does not apply for LLMs, and thus does not anticipate our attack. Finally, we investigate one line of possible defenses to our attack -- the addition of noise to the permutations \citep{morris2023textembeddingsrevealalmost}.

The main contributions of our paper are:

\begin{enumerate}
  \item We introduce a new attack on the hidden states of transformers to reverse permutations of them into the original prompts. We demonstrate the nearly-perfect performance of our attack on three different variants of hidden state permutation on Gemma 2 and Llama 3.1 across a range of layers. We discuss the key assumptions underlying the success of our attack, including the property of LLM hidden state \emph{non-collision}; the success of our attack provides strong evidence of this property being satisfied.
  \item We explain why the efficacy of our attack renders the schemes of \citet{zheng2024permllmprivateinferencelarge, yuan2024securetransformerinferenceprotocol, luo2024centaurbridgingimpossibletrinity} insecure; futher, we dissect the theoretical result based on distance correlation theory of \citet{yuan2024securetransformerinferenceprotocol, luo2024centaurbridgingimpossibletrinity}, and explain in detail why it does not anticipate our attack.
  \item Finally, we investigate a potential line of defenses to our attack, by the utilization of added noise.
\end{enumerate}

\section{Setup \& Threat Model}

We assume the setting of a user $U$ who wishes to perform inference with an LLM model $M$ on some input prompt $\boldsymbol{x}$, which can be considered as an ordered sequence of tokens $[x_1, x_2, ..., x_N]$. We denote the size of the hidden state of the LLM by $d$, and the sequence length by $N$. 

As the user $U$ does not have the resources to perform the inference themselves, they rely on a set of third-parties $P_1, P_2, ..., P_K$. We consider the setting where each of the parties behaves \emph{semi-honestly}, a common assumption of past works \citep{zheng2024permllmprivateinferencelarge, luo2024centaurbridgingimpossibletrinity, dong2023pumasecureinferencellama7b, yuan2024securetransformerinferenceprotocol}. Semi-honest parties will follow the defined protocol faithfully, but may exploit any information that they receive during the execution of the protocol to attempt to recover the user's data.

\begin{figure}[h!]
    \centering
    \includegraphics[width=0.9\linewidth, clip, trim=5 5 5 5]{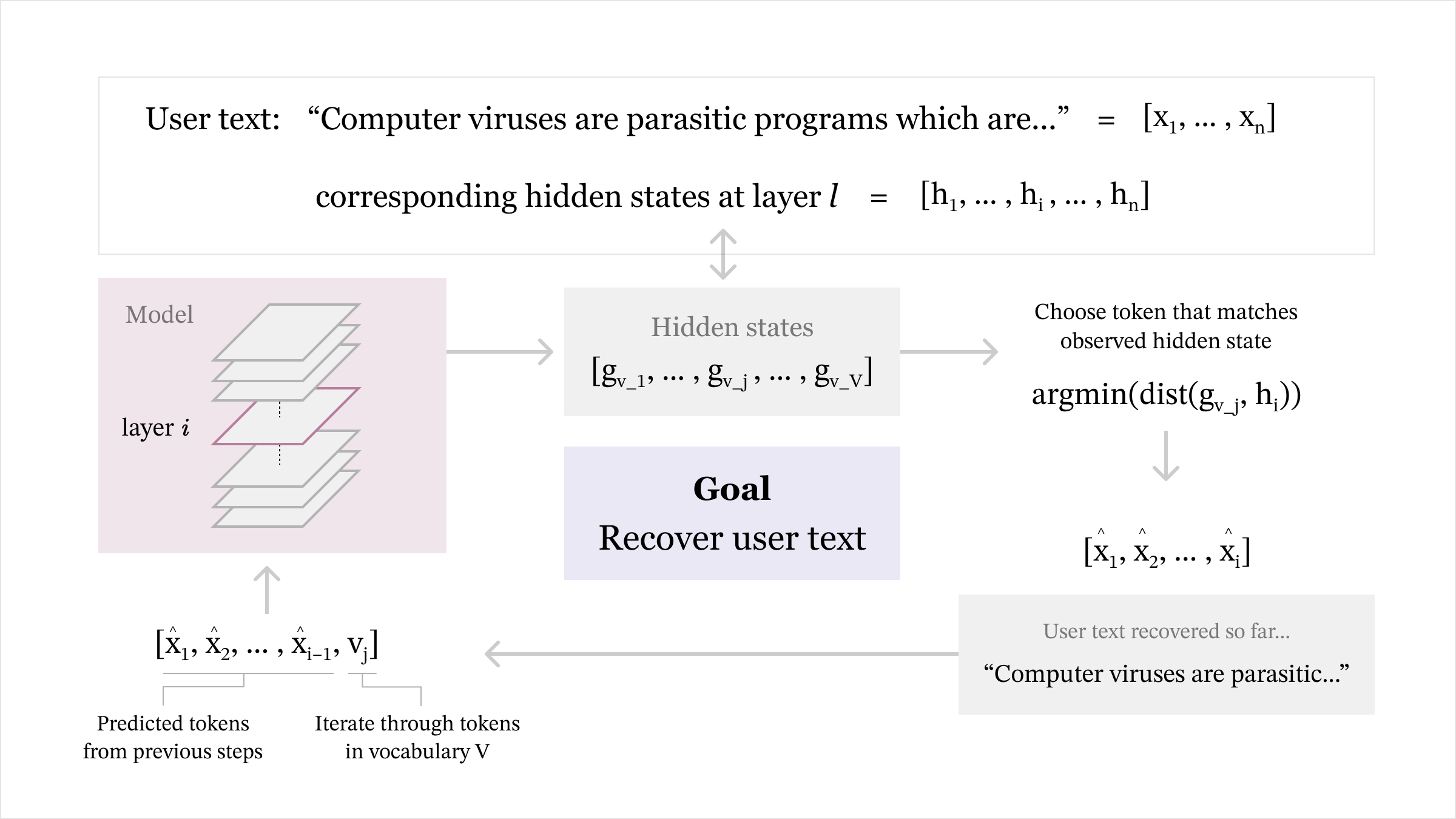}
    \caption{High-level representation of our attack to decode user text from LLM hidden states. This attack, and extensions of it, achieve nearly perfect decoding accuracy, even when the hidden states are permuted.}
    \label{fig:fig1}
\end{figure}

\section{Warmup: Unpermuted Hidden State Reversal}
\label{sec:hidden_reversal}

We begin by introducing our attack in the context of reversal of unpermuted hidden states -- a problem that has drawn prior attention in the literature in its own right (e.g. \citet{wan2024informationleakageembeddinglarge}). The key techniques we use to perform the attack in this setting will be the foundation that we further develop to perform reversal of permutations of hidden states later.

Consider the general case where one of the parties performing inference, $P_k$, receives an intermediate sequence of hidden states $\boldsymbol{h} = [h_1, h_2, ..., h_N]$ at some layer $l$ of the LLM $M$. Can the party $P_k$ reverse the hidden states $\boldsymbol{h}$ to the input sequence of tokens $\boldsymbol{x} = [x_1, x_2, ..., x_N]$ that produced $\boldsymbol{h}$?

\subsection{Informal Attack Description}
\label{subsec:vocab_matching_attack}

% Our proposed decoding scheme in the above setting leverages the causal ordering of decoder-only transformers, as well as the finite set of possibilities of input tokens. 

We outline our proposed attack below, and provide a visual depiction in \cref{fig:fig1}.

Our attack begins with a batched forward pass over all length-$1$ sequences $[v]$, where $v$ ranges over tokens in the vocabulary $\mathcal{V}$. From this, the adversary gets $V=|\mathcal{V}|$ candidate layer $l$ hidden states $\boldsymbol{h}(v) \in \mathbb{R}^{1 \times d}$. They set the first predicted input token $\widehat{x}_1$ to be the token $v$ for which $\boldsymbol{h}(v)$ matches the first hidden state $h_1$.

Next, the adversary performs a batched forward pass over all length-$2$ sequences $[\widehat{x}_1, v]$ with $v \in \mathcal{V}$, to get $V$ candidate layer $l$ hidden states $\boldsymbol{h}(\widehat{x}_1,v) \in \mathbb{R}^{2 \times d}$. Now, they set the second predicted input token $\widehat{x}_2$ to be the token $v$ where the second row of $\boldsymbol{h}(\widehat{x}_1,v)$ equals the second hidden state $h_2$.

In general, at the $n$th stage, using the first $n-1$ predicted input tokens $\widehat{x}_1,\ldots,\widehat{x}_{n-1}$, the adversary performs a forward pass over all length-$n$ sequences $[\widehat{x}_1,\ldots,\widehat{x}_{n-1},v]$ with $v \in \mathcal{V}$. They obtain $V$ candidate layer $l$ hidden states $\boldsymbol{h}(\widehat{x}_1,\ldots,\widehat{x}_{n-1},v) \in \mathbb{R}^{n \times d}$, and set the $n$th predicted input token $\widehat{x}_n$ to be the token $v$ where the $n$th (last) row of candidate states matches the $n$th hidden state $h_n$. Iterating over $n=1,\ldots,N$, the adversary sequentially obtains the predicted input sequence $\widehat{\boldsymbol{x}}$ from the layer $l$ hidden states $\boldsymbol{h}$.

Thus, although naively one may expect that an exact match of $\boldsymbol{h}$ would require exponential search (specifically, over all $V^N$ possible sequences of tokens $\boldsymbol{x}$), we see that this is reduced to a linear search; the total cost of this attack is $O(VN)$.

\subsection{Assumptions}
\label{subsec:assumptions_unpermuted}
The key assumptions necessary for our attack to succeed are:

\begin{enumerate}
    \item The forward pass performed over the vocabulary in the attack will match the forward pass that generated the given hiddens $\boldsymbol{h}$.
    \item Hidden states of LLMs are \emph{non-colliding}; that is, there is only one -- or at worst, a small number -- of matches between candidate tokens $v$ and hidden states $h_n$ at each step of the attack. If the average number of matches at each step is $M$, then the search space grows approximately as $M^N$, which is infeasible when $M$ or $N$ is large. As we shall see, this assumption turns out to be strongly satisfied in practice.
    \item The LLM has a \emph{unidirectional} attention structure. This is the case for decoder-only LLMs, which is the de-facto standard architecture for many current state-of-the-art LLMs.
    \item The model weights are available to the party $P_k$. Later, in the settings of \citet{yuan2024securetransformerinferenceprotocol, luo2024centaurbridgingimpossibletrinity}, we can relax this assumption.
\end{enumerate}

Assumption $1$ above is, however, not generally satisfied due to \textbf{non-determinism}.

\subsection{Non-Determinism}
\label{subsec:non-determinism}

% Prior work \citep{dong2023attentionneedpureattention} has demonstrated the rank-reducing effects of attention blocks, so it is plausible that the size of the subspace in latter layers in particular is too small to prevent large numbers of collisions.

\paragraph{Problem} In general, due to the non-associativity of floating-point operations \citep{villa2009effects}, we cannot expect that the attacker's candidate forward passes will exactly match the hidden states they already hold. Particularly in the GPU setting with parallel asynchronous thread execution and pooling without global synchronization, there can be considerable variation in the output \citep{shanmugavelu2024impactsfloatingpointnonassociativityreproducibility}. In addition, differences in hardware, random number seeds, environment variables and the state of initialized memory on the machine can all add to the variability, and these values may not be known to the attacker. Due to the presence of this reducible and irreducible noise, exact matching cannot be used successfully with this attack. 

\paragraph{Proposed Solution} To accommodate for this non-determinism, we loosen our matching requirements by computing the L1-distance between the last row of candidate hidden states and the given hidden state, and accept a match for a token $v$ if the distance is below some threshold $\boldsymbol{\epsilon}$. If no such match is found, we choose the token $v$ which gives minimal L1-distance.

However, by allowing an $\boldsymbol{\epsilon}$-ball for matching, we increase the possibility of collisions as described above in Assumption $2$. Is our attack still successful -- i.e., are LLM states sufficiently non-colliding -- even with this fuzzy matching? We find the answer is emphatically yes -- see \cref{subsec:experiments_hidden_unpermuted} below.

\begin{algorithm}[t]
\caption{Attack on Unpermuted LLM Hidden States}
\label{alg:vocab_matching_attack_optimized}
\begin{algorithmic}[1]
\INPUT Model $M$, layer $l$ hidden states $\boldsymbol{h} = [h_1, \ldots, h_N]$, vocabulary $\mathcal{V}$, proposal model $P$, L1-threshold $\boldsymbol{\epsilon}$%, KV-cache $C$
\OUTPUT Decoded token sequence $\widehat{\boldsymbol{x}} = [\widehat{x}_1, \widehat{x}_2, \ldots, \widehat{x}_N]$
\STATE Initialize empty sequence $\boldsymbol{\widehat{x}} \gets []$
%\STATE Initialize empty KV-cache $C \gets \{\}$
\FOR{$i = 1$ to $N$}
    \STATE $\mathcal{V}_{\text{ordered}} \gets \text{argsort}(P([\widehat{\boldsymbol{x}}, v] | \widehat{\boldsymbol{x}}))$ \COMMENT{Get ordered vocabulary from proposal model}
    \STATE $\text{min\_dist} \gets \infty$
    \STATE $\text{best\_match} \gets \text{None}$
    \FOR{$v \in \mathcal{V}_{\text{ordered}}$}
        \STATE $g \gets M_{\leq l}([\widehat{\boldsymbol{x}}, v])$ \COMMENT{Forward pass up to layer $l$} %with KV-caching}
        \STATE $\text{dist} \gets \| g - h_i|\|_1$ \COMMENT{Calculate L1 distance}
        \IF{$\text{dist} < \text{min\_dist}$}
            \STATE $\text{min\_dist} \gets \text{dist}$
            \STATE $\text{best\_match} \gets v$
        \ENDIF
        \IF{$\text{dist} < \boldsymbol{\epsilon}$} 
            \STATE $\widehat{x}_i \gets v$
            %\STATE $C \gets C'$ \COMMENT{Update KV-cache}
            \STATE break
        \ENDIF
    \ENDFOR
    \IF{$\text{min\_dist} \geq \boldsymbol{\epsilon}$}
        \STATE $\widehat{x}_i \gets \text{best\_match}$
        %\STATE $C \gets C'$ \COMMENT{Update KV-cache}
    \ENDIF
\ENDFOR
\STATE \textbf{return} $\widehat{\boldsymbol{x}}$
\end{algorithmic}
\end{algorithm}

\subsection{Efficiency}

We optimize runtime in practice using a \emph{proposal model} to provide a likelihood-based order of iteration through the vocabulary. We find that this modification reduces the average number of tokens searched through at each step from $V/2$ to $\sim 100$, resulting in a speedup of more than $1000\times$. In addition, we implement a novel variation of key-value-caching to further reduce the computational cost of our attack. Further details on these optimizations are given in \cref{appendix:attack_optimizations}. With these efficiency improvements, we reduce the decoding time of prompts of length 50 from many hours to typically around 2 minutes. 

\subsection{Formal Attack Description}

We now provide a formalized description of our attack, incorporating the modifications for efficiency and handling nondeterminism described above, in Algorithm \ref{alg:vocab_matching_attack_optimized}.

\subsection{Experiments}
\label{subsec:experiments_hidden_unpermuted}

We apply our attack on the hidden states of two state-of-the-art open-source LLMs, Gemma-2-2B-IT \citep{gemmateam2024gemma2improvingopen} and Llama-3.1-8B-Instruct \citep{grattafiori2024llama3herdmodels}. These models have different sizes (numbers of parameters), training methodologies, and architectures. We conduct testing on samples from the Fineweb-Edu dataset \citep{penedo2024finewebdatasetsdecantingweb}. The proposal model used is the same as the model being attacked. To ensure that there is no data leakage and that the dataset is unseen by the proposal model, we use the CC-MAIN-2024-10 data split, which postdates the models' training cutoff dates. We perform testing on hidden states taken from layers 1, 6, 11, 16, 21, and 26 of each model. For each layer of interest, we tune $\boldsymbol{\epsilon}$ by performing a ternary search on a small training set comprising 50 prompts taken from FineWeb, to determine the optimal L1-threshold under which predicted tokens are accepted as matches (note that an adversary could replicate this same tuning beforehand). We evaluate on 1000 held out prompts, and our results are shown in \cref{tab:decoding_acc_none}. 

\begin{table}[h!]
\centering
\caption{Percentage of (unpermuted) LLM hidden states that are perfectly decoded by our attack at different layers of Gemma-2-2B-IT and Llama-3.1-8B-Instruct, over 1000 samples of input prompts.}
\begin{tabular}{ccc}
\\ \toprule
\textbf{Layer} & \textbf{Gemma} & \textbf{Llama} \\
\midrule
1  &    100\%    &   100\%    \\
6  &    100\%    &   100\%     \\
11 &    100\%    &   100\%        \\
16 &    100\%    &   100\%     \\
21 &    100\%    &   99.9\%\\
26 &    100\%    &   99.7\%\\
\bottomrule
\end{tabular}
\label{tab:decoding_acc_none}
\end{table}

We find that nearly all evaluation samples are perfectly decoded. Accompanying $\boldsymbol{\epsilon}$ values are given in \cref{appendix:optimal_epsilons}. Due to computational constraints, each evaluation prompt was truncated to a maximum of 50 tokens; however, small-scale experiments with prompts exceeding 200 tokens demonstrated that our results generalize to longer prompt settings -- our attack still perfectly decodes nearly all hidden states into their corresponding tokens.

We further examine the rare cases where perfect decoding was not achieved. Errors mainly occur when prompts contain unexpected formatting characters, such as newline symbols (\textbackslash n) or hyphens (-). These artifacts sometimes cause the proposal model to favor an incorrect token, which is accepted early because its L1 error was below the $\boldsymbol{\epsilon}$ threshold. Although the correct token typically has a smaller error, the use of a proposal list to speed up the attack introduces this rare issue; tuning $\boldsymbol{\epsilon}$ using a larger training set may mitigate these errors, and using a full vocabulary search would completely avoid it, but at higher computational cost. Another source of error is prompts with accidental word repetitions (such as `the price price was high'), which occasionally disrupt the proposal model’s predictions in a similar way. Natural repetitions that are grammatically correct, however -- such as `he had had an operation' -- do not affect decoding accuracy.

% The success of our attack also allows us to conclude that LLM hidden states are highly distinct and non-colliding.

% \begin{table}[h]
% \centering
% \begin{tabular}{|l|c|c|l|}
% \hline
% \textbf{Model Size (num parameters)} & \textbf{Average Attack Time (s)} & \textbf{Vocabulary Size} & \textbf{Model Name} \\
% \hline
% 1B  & 49  & 128256 & Llama-3.2-1B-Instruct \\
% 2B  & 124 & 256000 & Gemma-2-2B-IT \\
% 8B  & 69  & 128256 & Llama-3.1-8B-Instruct \\
% 27B & 304 & 256000 & Gemma-2-27B-IT ($\epsilon = 30$) \\
% 27B & 124 & 256000 & Gemma-2-27B-IT ($\epsilon = 40$) \\
% \hline
% \end{tabular}
% \caption{Attack times and properties for various language models}
% \label{tab:model_attack_times}
% \end{table}

\section{Background: Permutation-Based Privacy-Preserving Schemes}
\label{sec:existing_work_permuted}

Recently, a number of works have proposed utilizing permutations to perform privacy-preserving inference of LLMs in a multi-party-computation (MPC) setup. We provide a description of three such schemes below.

\citet{zheng2024permllmprivateinferencelarge} introduces the \textbf{PermLLM} scheme. PermLLM permutes at the non-linear components of the LLMs in order to reveal them `safely' to one of the parties, and therefore avoid expensive iterated inter-party communication. The permutation is done on the attention logits before the softmax, at layer normalizations, and at the non-linear functions in the MLP block. The latter is a purely elementwise function, so the authors can do a full permutation across the $[N, d]$ elements, resulting in a permutation space of size $(Nd)!$. However, softmax and layer-norm are row-wise operations, so the permutation applied in this case is a (distinct) permutation to each of the columns, followed by a permutation of the $N$ rows, resulting in a permutation space of size $N!(d!)^N$.

\citet{yuan2024securetransformerinferenceprotocol} introduces the \textbf{STIP} scheme. In STIP, there are three parties: the model developer $P_1$, the model server $P_2$ (who carries out inference), and the user $P_3$. The goal of STIP is to have $P_2$ carry out inference on $P_3$'s input, protect $P_1$'s private model weights $\Theta$ from $P_2$ and $P_3$, and protect $P_3$'s private input data from $P_1$ and $P_2$. This is accomplished with random permutation in the hidden dimension. At initialization, $P_1$ sends random $d \times d$ permutation matrices $\pi,\pi_c$ to the user $P_3$, where $d$ is the token embedding dimension. They also randomly permute each weight matrix or vector in the row and/or column dimensions, to obtain the altered model weights $\Theta'$; these are given to the model server $P_2$, who cannot recover $\Theta$ from them. Then during inference, instead of sending their private input data $X \in \mathbb{R}^{N \times d}$, the user encrypts it with permutation $\pi$, i.e. they send $X \pi$. Then a standard transformer forward pass is carried out, but with the weights $\Theta$ (unknown to the model server $P_2$) replaced by permuted weights $\Theta'$. Finally, the results are sent to the user, who applies permutation $\pi_c$ to obtain the output of the inference. The STIP authors show through orthogonality of permutation matrices that the final output obtained is the same output as vanilla inference.

% permute both the model weights and the user prompt embeddings in the hidden $d$ dimension, and the entire inference process (on the next token) is then carried out by a single party.

\citet{luo2024centaurbridgingimpossibletrinity} introduces the \textbf{Centaur} scheme. Centaur follows the three-party threat model of STIP, and attempts to reconcile two problems. On the model weight privacy side, they aim to prevent exposure of the lookup table to the user. On the user privacy side, they wish to avoid exposing certain unpermuted intermediate results. For example, the authors observe that during the computation of attention, the calculation of $QK^T$ at each layer in STIP is insecure due to the $Q$ and $K$ permutations canceling. Therefore the authors apply the cryptographically-based technique of \textit{additive secret sharing} between the developer $P_1$ and server $P_2$ at most stages of self-attention, only requiring reconstruction of additive shares (by the developer) during nonlinearities. Although this resolves the previous two concerns, it is still the case that permutations of true layer $l$ hidden states are exposed to the model developer at nonlinearities.

All three schemes explicitly refer to the exponential difficulty of permutation reversal via brute-force attack, and therefore deem the revelation of permuted hidden states as secure. STIP and Centaur additionally make a theoretical claim of security based on distance correlation theory, which we address in \cref{sec:distcorr_theory}.

% Centaur applies ideas from both the above works. The proposed method permutes the model weights, utilizing additive secret-sharing for the linear layers, but relies on two-party permuted plaintext computation at the non-linearities (softmax, layer-norm and GeLU). Permutation is applied in the hidden $d$ dimension.

\section{Permuted Hidden State Reversal}
\label{sec:permuted_decoding}

We now consider the case where one of the parties performing inference receives a permutation of the intermediate sequence of hidden states $\boldsymbol{h}$ at some layer $l$ of the LLM $M$. We examine three different permutation types.

% We now propose a modification of our vocab-matching attack introduced in \cref{sec:hidden_reversal}, which breaks user input privacy for the above schemes in the open-weights setting. Extensions to the attack also break privacy in the closed-weights setting that \citet{yuan2024securetransformerinferenceprotocol} and \citet{luo2024centaurbridgingimpossibletrinity} originally consider: see \cref{appendix:closed_weights} for details. \citet{luo2024centaurbridgingimpossibletrinity} discuss theoretical considerations for why permutations should be statistically secure -- we discuss why these considerations do not anticipate or mitigate our attack in \cref{appendix:distance_correlation}.

\subsection{Sequence-Dim Permutation}
\label{subsec:sequence_dim_permutation}

First we consider a permutation in the \emph{sequence dimension}. Assume that permutation has been applied to layer $l$ hidden states $\boldsymbol{h} = [h_1, h_2, ..., h_N]$ such that:

\begin{equation*}
    \boldsymbol{h}_{\text{seq\_perm}} = [h_{\sigma(1)}, h_{\sigma(2)}, ..., h_{\sigma(N)}]
\end{equation*}

where $\sigma$ is a permutation of $[N]=\{1,2,\ldots,N\}$. 

\paragraph{Key Idea} The main insight we leverage to build our modified attack is that unidirectional attention imbues a positional marker on hidden state elements. That is, in any permuted sequence of LLM hidden states after at least one attention operation, there is exactly one `first' element that is a (non-linear) map of an embedding from the LLM's vocabulary and is \emph{not} a function of any other element. Similarly, there is exactly one `second' element that is a function of precisely the `first' element and another vocabulary embedding; and so on for the n'th element.

\paragraph{Attack Extension} We leverage the insight above to modify our attack from \cref{sec:hidden_reversal} as follows. At the first iteration of \cref{alg:vocab_matching_attack_optimized}, instead of finding the match to $h_1$, we now calculate the L1-distance to \emph{each} of the rows of $h$, and choose the vocabulary token $v$ which is within an L1-distance of $\boldsymbol{\epsilon}$ from \emph{any} row (if no match within $\boldsymbol{\epsilon}$ is found, we pick the minimum over all $v$ and rows of $h$). Let us assume that the match is made with the $j$th row of $h$ in the first iteration. In the second iteration, the $j$th row is removed from consideration; otherwise, the attack proceeds similarly to the first iteration, now considering all length-2 sequences $[\widehat{x}_1, v]$ with $v \in \mathcal{V}$ and matching against all remaining rows of $h$. This idea repeats for all $N$ iterations until the sequence is fully decoded.

The formal algorithm of our attack is given as \cref{alg:vocab_matching_attack_seq} in \cref{appendix:perm_algos}.

\paragraph{Assumptions} The main difference to our assumptions in the unpermuted setting (\cref{subsec:assumptions_unpermuted}) is that we now still require that hidden states of LLMs are non-colliding for a fixed position (or fixed number of prior tokens in the prompt), but are additionally non-colliding across \emph{all} possible positions (and number of prior tokens). That is, whilst previously the assumption required that $h_i$ was unique for a \emph{given} $i$, at any layer $l$ of the LLM, we now additionally require that $h_i$ is unique across \emph{all} position indices $i$.

% \paragraph{Key Insight} The Then, we modify the vocab-matching attack as follows. At the $n$th stage, we now choose the vocabulary token $v$ where the $n$th row of the corresponding candidate hidden state is within an L1-distance of $\boldsymbol{\epsilon}$ from \emph{any} row of $\boldsymbol{h}_{\text{seq\_perm}}$. Suppose this $\boldsymbol{\epsilon}$-ball match is made with the $i$th row $h_{\sigma(i)}$. We set the $n$th predicted input token $\widehat{x}_n$ to $v$, and remove $h_{\sigma(i)}$ from consideration for hidden state matching in \textit{all} future stages. Iterating over $n=1,\ldots,N$, we obtain the predicted input sequence $\widehat{\boldsymbol{x}}$ from sequence-permuted hidden states $\boldsymbol{h}_{\text{seq\_perm}}$.

% Compared to the vocab-matching attack, the opportunities for collision are now increased up to $N$-fold, as we match with up to $N$ rows of $\boldsymbol{h}$ rather than one. However, we again observe very few collisions in practice and are able to decode the vast majority of input prompts: see Table \ref{tab:combined_decoding_acc}.

% \begin{table}[h!]
% \centering
% \begin{tabular}{ccc}
% \toprule
% \textbf{Layer} & \textbf{Gemma} & \textbf{Llama} \\
% \midrule
% 1  &    100\%    &    99.7\%    \\
% 6  &    99.8\%    &   100\%     \\
% 11 &    100\%    &   100\%        \\
% 16 &    100\%    &   100\%     \\
% 21 &    99.8\%    &    100\%\\
% 26 &    99.8\%    &    100\%\\
% \bottomrule
% \end{tabular}
% \caption{The percentage of sequence permuted evaluation samples that were perfectly decoded}
% \label{tab:decoding_acc_s}
% \end{table}

\subsection{Hidden-Dim Permutation}
\label{subsec:hidden_dim_permutation}

Next we consider the case where permutation has been performed on the hidden dimension of $\boldsymbol{h}$ instead. That is, the party performing inference is now given:
\begin{equation*}
    \boldsymbol{h}_{\text{hidden\_perm}} = \left[\pi_1(h_1), \pi_2(h_2), ..., \pi_N(h_N)\right]
\end{equation*}
where each $\pi_i$ permutes elements of a $d$-dimensional vector.

\paragraph{Key Idea} In this setting, it is no longer possible to directly apply the L1-distance to find the nearest vocabulary token match. Instead, we use the \textbf{sorted L1-distance}, which individually sorts the two vectors to be compared and then computes their L1-distance. The effect of sorting is to map any two permutations in the hidden dimension to the same resultant vector. Note that the L1-distance is still required due to the existence of non-determinism discussed in \cref{subsec:non-determinism}.

\paragraph{Attack Extension} The modification to our attack from \cref{sec:hidden_reversal} is relatively straightforward; simply replacing the L1-distance at Step 8 of \cref{alg:vocab_matching_attack_optimized} with the sorted-L1 distance as described above instead. The formal algorithm of our attack is given as \cref{alg:vocab_matching_attack_hidden} in \cref{appendix:perm_algos}.

\paragraph{Assumptions} Now the main difference to our assumptions in the unpermuted setting (\cref{subsec:assumptions_unpermuted}) is that we additionally require that LLM hidden states are non-colliding even when they are sorted (in the hidden dimension). In essence, this assumption is equivalent to the assertion that permuting LLM hidden states in the hidden dimension offers no obfuscation at all -- for a fixed sequence position $i$, they are still uniquely identifiable. Indeed, due to the existence of non-determinism, we actually require an even stronger property -- since the noise exists \emph{before} the sorting is done, even the same vector with two different noises applied can end up with a different ordering after sorting. Therefore, we also require that LLM hidden states are robust to the noise introduced by non-determinism such that if sorting results in a different ordering to the correct token, that ordering is still the closest vector by L1-distance versus all other noisy sortings of tokens in the vocabulary.
% Again, the existence of noise may appear to be a significant obstacle to this approach. However, we find that even this relatively simple matching approach is robust enough to noise to achieve nearly perfect decoding. Our results are shown in \cref{tab:combined_decoding_acc}.

% \begin{table}[h!]
% \centering
% \begin{tabular}{ccc}
% \toprule
% \textbf{Layer} & \textbf{Gemma} & \textbf{Llama} \\
% \midrule
% 1  &    100\%    &   100\%      \\
% 6  &    100\%    &   98.5\%     \\
% 11 &    100\%    &   99.2\%     \\
% 16 &    99.9\%   &   99.4\%     \\
% 21 &    98.2\%   &   98.9\%     \\
% 26 &    98\%     &   98.2\%     \\
% \bottomrule
% \end{tabular}
% \caption{The percentage of hidden dimension permuted evaluation samples that were perfectly decoded}
% \label{tab:decoding_acc_d}
% \end{table}

\subsection{Factorized-2D Permutation}
\label{subsec:factorized_2d_permutation}

We now consider the case of a factorized two-dimensional permutation as used in \citet{zheng2024permllmprivateinferencelarge}, where a hidden-dimension permutation is applied to each hidden state, and then these resulting states are shuffled in the sequence dimension. The adversary now has:
\begin{equation*}
    \boldsymbol{h}_{\text{fact\_perm}} = [\pi_1(h_{\sigma(1)}), \pi_2(h_{\sigma(2)}), ..., \pi_N(h_{\sigma(N)})]
\end{equation*}
where $\sigma$ is a permutation of $[N]$ and each $\pi_i$ permutes a $d$-dimensional vector. 

\paragraph{Key Idea} For this setting, we combine ideas from both the sequence-dim extension and the hidden-dim extension above. The principles of both extensions remain true -- that at the $n$th matching stage, there is a single element that is a function only of the previous $n - 1$ elements and some token $v$ in the vocabulary; and the use of the sorted-L1 matching function to be able to compare two different hidden-dimension permutations of the same vector.

\paragraph{Attack Extension} The formal algorithm of our attack extension to the factorized-2D setting is given as \cref{alg:vocab_matching_attack_2D} in \cref{appendix:perm_algos}.

\paragraph{Assumptions} The factorized-2D permutation setting requires the strongest uniqueness assumption among the three settings. Specifically, we require that hidden states are non-colliding across all possible sequence positions and numbers of prior tokens, even after sorting the elements of each hidden state vector. That is, at any layer $l$ of the LLM, the sorted hidden state must be uniquely identifiable across all positions $i$, even in the presence of the non-deterministic noise prior to the sorting operation being applied.

% \begin{table}[h!]
% \centering
% \begin{tabular}{ccc}
% \toprule
% \textbf{Layer} & \textbf{Gemma} & \textbf{Llama} \\
% \midrule
% 1  &    99.9\%    &   98.4\%    \\
% 6  &    99.5\%    &   97.8\%     \\
% 11 &    99.5\%    &   98.9\%        \\
% 16 &    99.2\%    &   98.8\%     \\
% 21 &    99.1\%    &   98\%\\
% 26 &    99.0\%    &   97.6\%\\
% \bottomrule
% \end{tabular}
% \caption{The percentage of factorized 2D permuted evaluation samples that were perfectly decoded}
% \label{tab:decoding_acc_sd}
% \end{table}

\subsection{Experiments}
\label{subsec:permutation_experiments}

We now measure the efficacy of our attack on permutations of the hidden states of Gemma-2-2B-IT and Llama-3.1-8B-Instruct. We again take samples from the Fineweb-Edu dataset's CC-MAIN-2024-10 data split, post-dating the models' training cutoff dates. We apply sequence dimension, hidden dimension, and 2D permutation to each of the hidden states as described above. We again test across a range of model layers, and tune $\boldsymbol{\epsilon}$ in each setting by performing a ternary search on a small training set comprising 50 prompts. We evaluate on 1000 held out prompts in each setting. Our results are shown in \cref{tab:combined_decoding_acc}. 

% \begin{table}[h!]
% \centering
% \caption{The percentage of evaluation samples that were perfectly decoded under sequence-dim, hidden-dim, and factorized 2D permutations, for Gemma-2-2B-IT and Llama-3.1-8B-Instruct.}
% % \scriptsize
% \begin{tabular}{ccccccccc}
% \\ \toprule
% \textbf{Layer} & \multicolumn{2}{c}{\textbf{Sequence-Dim}} & \multicolumn{2}{c}{\textbf{Hidden-Dim}} & \multicolumn{2}{c}{\textbf{Factorized-2D}} \\ 
% \cmidrule(lr){2-3} \cmidrule(lr){4-5} \cmidrule(lr){6-7}
%  & \textbf{Gemma} & \textbf{Llama} & \textbf{Gemma} & \textbf{Llama} & \textbf{Gemma} & \textbf{Llama} \\ 
% \midrule
% 1  & 100\%    & 99.7\%  & 100\%   & 100\%   & 99.9\%  & 98.4\%  \\
% 6  & 99.8\%   & 100\%   & 100\%   & 98.5\%  & 99.5\%  & 97.8\%  \\
% 11 & 100\%    & 100\%   & 100\%   & 99.2\%  & 99.5\%  & 98.9\%  \\
% 16 & 100\%    & 100\%   & 99.9\%  & 99.4\%  & 99.2\%  & 98.8\%  \\
% 21 & 99.8\%   & 100\%   & 98.2\%  & 98.9\%  & 99.1\%  & 98.0\%  \\
% 26 & 99.8\%   & 100\%   & 98.0\%  & 98.2\%  & 99.0\%  & 97.6\%  \\
% \bottomrule
% \end{tabular}
% \label{tab:combined_decoding_acc}
% \end{table}

\begin{table}[h!]
\centering
\caption{The percentage of evaluation samples that were perfectly decoded under sequence-dim, hidden-dim, and factorized 2D permutations, for Gemma-2-2B-IT and Llama-3.1-8B-Instruct.}
% \scriptsize
\begin{tabular}{ccc}
\\ \toprule
\textbf{Layer} & \multicolumn{2}{c}{\textbf{Factorized-2D}} \\ 
\cmidrule(lr){2-3}
 & \textbf{Gemma} & \textbf{Llama} \\ 
\midrule
1  & 99.9\%  & 98.4\%  \\
6  & 99.5\%  & 97.8\%  \\
11 & 99.5\%  & 98.9\%  \\
16 & 99.2\%  & 98.8\%  \\
21 & 99.1\%  & 98.0\%  \\
26 & 99.0\%  & 97.6\%  \\
\bottomrule
\end{tabular}
\label{tab:combined_decoding_acc}
\end{table}

As can be seen, our attack remains highly effective under all of the permutation types described above, and across all layer choices. Sequence-dimension permutation in particular is decoded at essentially a 100\% success rate. The success of our attack against hidden-dimension permutation is also above 99\% for earlier layers, though it does drop slightly in the later layers of both Gemma and Llama. We theorized above that factorized-2D permutation decoding requires the strongest conditions on LLM hidden state non-collision, and this is borne out by the slightly lower decoding results for this setting than the other two permutation types. However, the attack maintains a 99\% decoding rate across all layers for Gemma, with a slightly reduced success rate for Llama. Moreover, we emphasize that our metric counts the number of \emph{perfect} decodings, and even in the cases where this was not obtained, we observed significant partial decoding of the original prompt. Our results largely support our assumptions on the non-colliding properties of LLM hidden states described in each of the sections above.

\section{Implications For Permutation-Based Privacy-Preserving Schemes}

We now describe the implications of the efficacy of our family of attacks for the schemes described earlier in \cref{sec:existing_work_permuted}.

\paragraph{PermLLM} Recall that PermLLM reveals the permuted hidden states at the non-linearities to the parties performing inference; and that the hiddens at the softmax and layer-norm non-linearities, in particular, undergo factorized-2D permutation as they are row-wise operations. Therefore, any party that receives the hidden states at these non-linearities, at any layer, can directly apply the attack described in \cref{subsec:factorized_2d_permutation}, with the very high success rates demonstrated in \cref{subsec:permutation_experiments}.

\paragraph{STIP} Recall that in STIP, party $P_2$ carries out inference using a model with permuted weights $\Theta'$, on a permutation of the input, $X\pi$, in the hidden dimension. Apart from an additional detail regarding access to the embedding layer, which we expand on below in \cref{subsec:private_embedding}, this is analogous to the hidden-dimension permutation setting. A forward pass from the altered transformer model with weights $\Theta'$ up to layer $l$ will allow $P_2$ to recover hidden-dimension-permuted layer $l$ hidden states, and apply the attack from \cref{subsec:hidden_dim_permutation} to recover the input.

\paragraph{Centaur} Centaur operates similarly to STIP from the perspective of our attack; at the non-linearities, hidden-dimension-permuted hidden states are revealed to the parties performing inference; and party $P_2$ has access to the permuted weights $\Theta'$. Therefore, the attack of \cref{subsec:hidden_dim_permutation} can also be used on Centaur.

\subsection{Private Embedding Layer}
\label{subsec:private_embedding}

In both STIP and Centaur, the party which performs inference, $P_2$ has access to the entire set of permuted model weights $\Theta'$ -- \emph{except} for the token embedding layer, which is not revealed to $P_2$. This lookup table is instead only revealed to the user, $P_3$, who embeds their prompt using this, permutes the embeddings in the hidden dimension, and then sends them to $P_2$ for inference. As such, without direct access to the possible token embeddings, the adversary cannot immediately carry out full candidate forward passes, a crucial element of our attacks. However, this is straightforwardly circumventable; in many modern LLM families, the embedding matrix is simply the tranpose of the language-modeling head, whose permutation is known to $P_2$. Even if this is not the case, $P_2$ may build their own `vocabulary' of input embeddings from observing repeated inference requests and then carry out our attack. We give further details on both of these in \cref{appendix:private_embedding}.

\section{Distance Correlation Does Not Guarantee Permutation Security}
\label{sec:distcorr_theory}

We now contextualize statistical arguments on the security of permuted hidden states. In particular, we clarify why they do not anticipate our attack.

Both STIP and Centaur rely on results from distance correlation theory \citep{szekely2007measuring} to support their arguments on the security of permuted hidden states. Citing \citet{zheng2022towards}, both papers quote the following result:

\begin{align}
\mathbb{E}_{\pi, W_A \in \mathbb{Z}^{d \times d}} \left[ \mathrm{Discorr}(x, x W_A \pi) \right]
\leq 
\mathbb{E}_{W_B \in \mathbb{Z}^{d \times 1}} \left[ \mathrm{Discorr}(x, x W_B) \right].
\label{eqn:discorr}
\end{align}

where Discorr is the distance correlation function and $x \in \mathbb{R}^d$ is the input vector chosen from a data distribution. Here, the expectations are taken over $W_A$ and $W_B$ sampled from standard random normal distributions and $\pi$ sampled uniformly over all $d!$ permutation matrices. In essence, this result demonstrates that the expected distance correlation between any vector and the same vector with a random permuted (dimensionality-preserving) linear map applied, is less than the expected distance correlation between the vector and the same vector with a 1-dimensional compressing linear map applied. The authors claim that therefore, permuted LLM hidden states retain less information about the input embeddings than a 1-D projection.

There are at least three reasons why this result cannot be used to make strong guarantees on the security of their schemes, which we outline in the following subsections.

\subsection{Reconstruction From Random 1D Projections Is Feasible}

The authors assert that reconstructing inputs after a random $1$-dimensional linear projection is difficult. However, there is no theoretical reason that this should be the case, especially for such projections of LLM hidden states.

    We can make this statement precise as follows. Our attack is able to successfully reverse LLM hidden states with L1-distance matching as demonstrated in \cref{subsec:experiments_hidden_unpermuted} and \cref{subsec:permutation_experiments}. Assuming that two vectors are non-colliding with respect to L1-distance, we can ensure random 1D projections of these two vectors are also non-colliding with high probability. 

    \begin{theorem}
    \label{theorem:1dproj}
       Let $k>0$. Suppose random weights $\bm{w} \in \mathbb{R}^d$ are drawn from a $d$-variate spherically symmetric distribution $\mathcal{D}$. Then any $\bm{x},\bm{y} \in \mathbb{R}^d$, we have the absolute difference of $\bm{w}$-weighted sums of $\bm{x}$ and $\bm{y}$ exceeds the L1 distance between $\bm{x}$ and $\bm{y}$ by a factor $\geq k$, meaning
        \begin{equation}\label{eqn:random_l1}\left|\sum_{i=1}^d w_i x_i - \sum_{i=1}^d w_i y_i\right| \geq k \sum_{i=1}^d |x_i-y_i|,\end{equation}
        with probability $\geq P_{\bm{\gamma} \sim \mathcal{D}} (|\gamma_1| \geq k\sqrt{d})$.
    \end{theorem}

    \begin{proof}
        See \cref{appendix:proof_1dproj}.
    \end{proof}

    Although the above holds over all spherically symmetric distributions, we can obtain an exact bound above by setting $\mathcal{D}$ to a multivariate Gaussian. That is, for $\bm{w}=(w_1,\ldots,w_d)$, we i.i.d. sample each $w_i \sim \mathcal{N}(0,\sigma^2)$. Then the lower bound in the theorem is $P_{\bm{\gamma} \sim \mathcal{D}} (|\gamma_1| \geq k\sqrt{d}) = P_{\bm{\gamma} \sim \mathcal{N}(0,\sigma)} (|\gamma| \geq k\sqrt{d}) = 2- 2\Phi(k\sqrt{d}/\sigma)$, where $\Phi$ is the normal CDF. With sufficiently large $\sigma$ or small $k$, we can make this lower bound approach $2 - 2\Phi(0) = 1$. For instance, for $d=4096$ in Llama-3.1-8B-Instruct, if we sample weights with $\sigma=1$ (as is done by \citet{zheng2022towards} in the statement of \cref{eqn:discorr}), setting $k=1/64$ gives a lower bound of $2-2\Phi(1) \approx 32\%$, and setting $k=1/32$ gives a lower bound of $2-2\Phi(2) \approx 5\%$. To increase $k$ (for a stronger guarantee of non-collision of the weighted sums) while maintaining the probability lower bound, one must proportionally increase the standard deviation $\sigma$ of the random weights.
    
    It is therefore plausible that even with access to random 1D linear projections of LLM hidden states, our attack would be successful. Further work should experimentally verify the efficacy of our attack with randomly-weighted sums, in the presence of non-determinism and other practical implementation considerations.

\subsection{Distance Correlation Misaligns With Reconstructibility}

To measure privacy leakage, \citet{zheng2022towards} use expected distance correlation. They justify their choice by noting distance correlation is a well-known statistical metric, which represents structural similarity between datasets and is straightforward to estimate. However, as we now show, distance correlation is not a universal measure of how reversible one random variable is from another. To demonstrate this shortcoming, we introduce the notion of `$\delta$-reconstructibility', which captures the ability to recover one variable from another variable up to a given absolute threshold. We define it formally as:

\begin{definition}
Let $X,Y$ be random variables. We say that $(X,Y)$ is \textbf{$\delta$-reconstructible} if there exists a function $f(Y)$ such that $|X-f(Y)| \leq \delta$ almost always. 
\end{definition}

This notion of $\delta$-reconstructibility is directly tied to privacy in our setting, as the non-determinism described in \cref{subsec:non-determinism} forces us to choose the candidate token within a given absolute threshold. We now show that $\delta$-reconstructibility does not align with distance correlation: there are $\delta$-reconstructible pairs with a lower distance correlation than non-$\delta$-reconstructible pairs.

\begin{theorem}
    For any $\delta>0$, there exist random variables $W,X,Y,Z$ such that $\text{Discorr}(W,X) > \text{Discorr}(Y,Z)$, the pair $(W,X)$ is not $\delta$-reconstructible, and the pair $(Y,Z)$ is $\delta$-reconstructible. 
    \label{theorem:discorr}
\end{theorem}

\begin{proof}
    See \cref{appendix:proof_discorr}.
\end{proof}

Additionally, we observe that \cref{eqn:discorr} involves an expectation of distance correlation over random linear maps and permutations. Therefore, it is possible that there are particular linear weights and permutations where the distance correlation with a randomly permuted linear projection is smaller than the distance correlation with a random 1D linear projection. Therefore, \cref{eqn:discorr} cannot be applied to make universal claims about reconstructibility across different models and permutations.

\subsection{Transformers Have Token Interdependence}

    Even taking \cref{eqn:discorr} at face value, it is still questionable how it proves security for \textit{transformer models}. Linear projections are only one component of these architectures: a formal security guarantee should incorporate the other modules in a transformer, especially self-attention, in which tokens are not processed independently. In particular, this means a valid result should be proved over a distribution over full $N \times d$ inputs, rather than a distribution of $1 \times d$ embeddings as in \cref{eqn:discorr}. In fact, the unidirectional nature of decoder-only LLMs through self-attention is a key assumption that enables the vocabulary-matching attack to succeed (\cref{subsec:assumptions_unpermuted}). Thus, the distance correlation result, which ignores this dependence, fails to anticipate such an attack.  
    
    %Furthermore, the result does not anticipate discrete information leakage from the lookup table, which is also key to vocabulary-matching in handling token dependence. Although STIP and Centaur attempted to make the tokenizer and embedding layer private to avoid such risks, we showed in \cref{subsec:private_embedding} that repeated inference calls and a substitution cipher can still be used to decode the input in this setting.

\section{Investigation of Possible Defenses}
\label{sec:noised_decoding}

Having demonstrated the efficacy of our attack family in decoding permuted hidden states of LLMs, we now investigate potential defensive approaches that still permit the general idea of revelation of permuted plaintext to parties, without incurring severe information leakage. We focus our investigation on defensive measures that aim to disrupt Assumption 2 of \cref{subsec:assumptions_unpermuted} by the use of various \textbf{noising} approaches.

% \cref{sec:permuted_decoding} shows that modifications to our attack can successfully decode any sequence-dimension, hidden-dimension, and factorized-2D permutation of the hidden states. We now examine the efficacy of our attack on alternative methods of defense that modify the hidden states directly -- such as by adding noise, or by quantizing the model to a lower precision. We find that generally, these methods are still not sufficient to defend against our attack.

We investigate the following methods of modification to the permuted LLM hidden states:

\begin{itemize}
    \item Adding diagonal Gaussian noise with mean 0 and standard deviation $\sigma$ to each hidden dimension in the input embeddings, as proposed in \citet{morris2023textembeddingsrevealalmost}.
    \item Inserting a randomly generated embedding as a prefix to the original sequence. This has the effect of modifying the subsequent hidden states via self-attention.
    \item Quantization of the model.
\end{itemize}

Clearly, with a sufficiently high degree of noise, decoding can be made impossible. However, high noise will also likely disrupt LLM performance. Therefore, the crux of any such defense is based on the delicate balancing act of ensuring security against our attack, whilst still maintaining downstream model performance. 

\subsection{Experiments}

We apply each of the above noising methods on Gemma-2-2B-IT. For diagonal Gaussian noise, we test with $\sigma=0.1, 0.01$. For the random embedding prefix, we generate the embedding from a Gaussian with means and standard deviations of each hidden dimension set to the average over the token vocabulary $\mathcal{V}$. For quantization, we test with reduction of the model from its original 16-bit to 8-bit and 4-bit, using the bitsandbytes library \citep{bitsandbytes}. We apply each of the above methods to all the permutation types described in \cref{sec:permuted_decoding}, as well as the unpermuted hidden states. Our choice of dataset, number of evaluation samples, and method of choosing $\boldsymbol{\epsilon}$ is the same as in \cref{subsec:experiments_hidden_unpermuted} and \cref{subsec:permutation_experiments}. As perfect decoding is less commonly achieved with the addition of noise, we now report the ROUGE-L score between the decoded reconstruction and the original prompt to measure decoding quality.  We conduct testing again over layers 1, 6, 11, 16, 21 and 26, but report only the highest ROUGE-L, as this can be considered the weakest attack point.

 To measure the downstream impact of the noising methods, we utilize LiveBench \citep{white2024livebenchchallengingcontaminationfreellm}, a benchmark that tests across multiple different components of LLM performance, such as language, reasoning and math. Our results are given in \cref{tab:noised_decoding} below. A full breakdown of the LiveBench scores by category and the ROUGE-L scores by layer of each of the above methods and permutation types is given in \cref{appendix:noising_method}. 

\begin{table}[h!]
\centering
\caption{ROUGE-L reconstruction scores across 1000 evaluation samples for various noising methods and permutation types on Gemma-2-2B-IT. The `Downstream Performance' column is the normalized score on LiveBench \citep{white2024livebenchchallengingcontaminationfreellm}, a benchmark that tests broad components of LLM performance such as math, reasoning and language. Note that LiveBench scores carry some variability, and so the baseline, Gaussian with standard deviation $0.01$, and random embedding prefix methods are all within noise in performance.} 
\resizebox{\textwidth}{!}{
\begin{tabular}{ccccc|c}
\\ \toprule
\textbf{Method} & \textbf{Unpermuted} & \textbf{Sequence Perm} & \textbf{Hidden Perm} & \textbf{Factorized 2D} & \textbf{Downstream Performance} \\
\midrule
Baseline (no noise) & 1.00 & 1.00 & 1.00 & 1.00 & 100.0\% \\
\cmidrule(lr){1-6}
Gaussian, $\sigma=0.01$ & 0.93 & 0.07 & 0.07 & 0.07 & 101.4\% \\
Gaussian, $\sigma=0.1$ & 0.91 & 0.01 & 0.01 & 0.01 & 5.8\%  \\
Random emb. prefix         & 0.93 & 0.17 & 0.19 & 0.19 & 102.9\% \\
8-bit quantization         & 0.89 & 0.86 & 0.75 & 0.73 & 97.6\% \\ 
4-bit quantization         & 0.88 & 0.84 & 0.83 & 0.71 & 92.2\% \\ 
\bottomrule
\end{tabular}
}
\label{tab:noised_decoding}
\end{table}

We see that unpermuted hidden states are still highly decodeable via our attack under all methods tested -- the ROUGE-L scores are above 0.8 in all cases, indicating significant similarity with the original text. Remarkably, even 4-bit quantization is not sufficient to introduce enough collisions to significantly mitigate our attack. We find that the combination of permutation and Gaussian noise with standard deviation $0.01$ appears largely secure, with ROUGE scores below 0.1, and maintains downstream performance, and thus may represent a potential solution to the insecurity of STIP and Centaur. However, this result is only necessary for security, and not sufficient; it is possible that extensions of our attack family can succeed even in this setting. We leave further investigation of this to future work.

\section{Related Work}
\label{sec:related_work}

Several existing works have investigated the reversibility of LLM embeddings into the original sentence inputs \citep{song2020informationleakage, morris2023textembeddingsrevealalmost, li2023sentenceembeddingleaksinformation, kugler2024invbert} with relatively good decoding performance. Different from our setting, these focus on reversal of a single vector $\boldsymbol{e} = \phi(\boldsymbol{x}) \in \mathbb{R}^d$, where $\phi$ is an embedding model that returns a single fixed-size vector from an $N$-token input $\boldsymbol{x} = [x_1, x_2, ..., x_N]$. In our paper, we are instead concerned with the reversibility of full intermediate states $[h_1,h_2,\ldots,h_N] \in \mathbb{R}^{N \times d}$ of an LLM.

The closest two previous works on reversibility in our setting are those of \citet{wan2024informationleakageembeddinglarge} and \citet{morris2023languagemodelinversion}. The former work focuses on reversal of hidden states in general, whilst the latter is particularly focused on logit output distribution reversal. In both papers, the authors use a learnt transformer-based network to reverse the sequence of hidden states into the original token inputs. Experiments are conducted on two decoder-based models, Llama-2-7B and ChatGLM-6B. Average F1 scores of approximately $60\%$ are achieved across a range of datasets in \citet{wan2024informationleakageembeddinglarge} on hidden states near the last layers of the models, and scores around $75\%$ are achieved for logit reversal in \citet{morris2023languagemodelinversion}. Importantly, the latter paper does not assume any access by the adversary to model weights, whilst the former explicitly denotes the case of a model provider performing inference on user provided embeddings, and so is more analogous to our setting.

\citet{petrov2024dager} propose an attack that shares some elements with ours below -- especially, exploitation of the unidirectional nature of decoder-based LLMs, as well as the finite and discrete space of LLMs' vocabularies. However, they are concerned primarily with the setting of gradient reversal into original inputs in the federated \textit{training} setting -- different from our focus on private \textit{inference}. Furthermore, their method relies on full-rank properties of the gradients, which are not always satisfied (e.g. when the prompts are longer than the hidden dimension size). By contrast, our method does not have any such restrictions. 

To the best of our knowledge, no existing work specifically attempts to, or succeeds at, reversing permutations of LLM hidden states.

% Finally, their method requires knowledge of the LLM gradients with respect to \textit{first layer} weights, but our attack only needs \textit{some layer} of LLM hidden states to work.

\section{Conclusion \& Future Work}

We have introduced a new attack for decoding LLM hidden states into their original user text. We have demonstrated the efficacy of this attack for reversal of LLM hidden states into their original prompts. We then proposed extensions of this attack that we have shown are capable of nearly-perfect reversal of various types of permutations of LLM hidden states, compromising the security of three previously proposed private-inference schemes. We also deconstructed previous assertions of security based on misapplications of distance correlation theory. Finally, we have investigated a potential line of defenses to our attack -- the addition of noise to the LLM hidden states.

There are several promising future directions of research that build on our contributions in this work. We have not yet demonstrated a successful attack against \emph{unrestricted} permutations of hidden states, i.e. where any element of the $N \times d$ matrix of hiddens can be moved to any column or row index without restriction. Although this is not necessary to break the security of the schemes we analyze in this paper, such a scheme may be proposed in the future -- for example, where only the hiddens at the elementwise non-linearities are revealed as permuted plaintext. Additionally, 
further work should investigate the security of combining noise and permutations against our attack, as we propose in \cref{sec:noised_decoding}.

\bibliography{ref}
\bibliographystyle{plainnat}

\appendix
\section{Attack Optimizations}
\label{appendix:attack_optimizations}

\paragraph{Proposal Model} Although the cost of the attack outlined in \cref{sec:hidden_reversal} is linear in $V$, the size of vocabularies can be quite large in practice. For example, Gemma-2-2B-IT has a vocabulary size of $256000$. Therefore we seek to optimize this by introducing a \emph{proposal model}. The purpose of the proposal model is to provide a suggested ordering over the vocabulary, rather than iterate through it in an arbitrary order. It does so by taking in the token sequence that has been partially decoded so far and producing the next-token logits. We then search through the next-token logits in decreasing order of probability. In practice, we find that this modification reduces the expected number of tokens searched through at each step from $V/2$ to approximately $100$, thus representing a constant factor speedup of more than $1000\times$.

\paragraph{KV-Caching} Additionally, we implement a novel variation of key-value-caching (KV-caching) to reduce the computational time of our attack. Note that at the $n$th stage of the decoding, we are performing a $V$-batched forward pass on $[\widehat{x}_1, \widehat{x}_2, ..., \widehat{x}_{n-1},v]$ over $v \in \mathcal{V}$, where $\widehat{x}_1,\widehat{x}_2,\ldots,\widehat{x}_{n-1}$ are the tokens that we have already decoded. As this forward pass needs to be repeated many times for different $v$ but the same $\widehat{x}_i$, we cache the keys and values associated to the $\widehat{x}_i$ and reuse them across all forward passes. This is different from standard KV-caching, which stores the keys and values for generation over a single sequence: here, we reuse keys and values across many sequences. In practice, this optimized caching gives a significant speedup in our attack: across $10$ evaluation prompts from FineWeb, the average caching speedup was around $20 \times$, with speedups for all prompts in the range $15$-$30\times$.

\section{Optimal $\boldsymbol{\epsilon}$ for Decoding}
\label{appendix:optimal_epsilons}

We report the full set of optimal $\boldsymbol{\epsilon}$ thresholds in decoding, for each permutation type below. We observe that generally, the optimal $\boldsymbol{\epsilon}$ increases in later layers across all permutation types -- which may be due to the effect of the reducible and irreducible noise we mentioned in \cref{sec:hidden_reversal} taking up a larger subspace volume as it propagates to deeper layers. We also observe that Llama tends to have much lower $\boldsymbol{\epsilon}$ values in general. Further investigation of these interesting trends and their implications for the properties and structure of LLM hidden states is left to future work.

There is also an interesting distinction between Gemma and Llama to note in this trend: the optimal $\boldsymbol{\epsilon}$ for the last hidden layer (26) in Gemma decreases by nearly $2\times$ from the previous tested layer (21) outside of the no permutation case. But the opposite is the case for Llama: it increases by more than $2 \times$ at the last layer (32) from the previous tested layer (26) for all but the no permutation case. Both Gemma and Llama have slightly decreased $\boldsymbol{\epsilon}$ at the last layer in the no permutation case. Investigating the reason for decreasing versus increasing $\boldsymbol{\epsilon}$-ball collisions in the last few layers, based on distinctions in the architecture or weights of models like Gemma and Llama, and the type of permutation applied, is an interesting direction for future work.

\FloatBarrier
\begin{table}[h!]
\centering
\caption{Optimal $\boldsymbol{\epsilon}$ thresholds for hidden state reversal with no permutation, over various Gemma-2-2B-IT and Llama-3.1-8B-Instruct layers.}
\begin{tabular}{ccc}
\toprule
\textbf{Layer} & \textbf{Gemma} & \textbf{Llama} \\
\midrule
1  &    22.0    &  0.6      \\
6  &    70.0    &  7.1     \\
11 &    204.0    &  18.3      \\
16 &    293.0    &  29.0      \\
21 &    400.0    &  76.0      \\
26 &    318.0    &  156.0      \\
32 &    ---      &  150.0    \\
\bottomrule
\end{tabular}
\label{tab:optimal_eps_none}
\end{table}

\begin{table}[h!]
\centering
\caption{Optimal $\boldsymbol{\epsilon}$ thresholds for hidden state reversal with sequence dimension permutation, over various Gemma-2-2B-IT and Llama-3.1-8B-Instruct layers.}
\begin{tabular}{ccc}
\toprule
\textbf{Layer} & \textbf{Gemma} & \textbf{Llama} \\
\midrule
1  &    12.8   &  1.4     \\
6  &    72.6     &  3.3     \\
11 &    229.0    &  7.4      \\
16 &    301.0    &  7.4      \\
21 &    385.0    &  26.6      \\
26 &    220.0    &  29.6      \\
32 &    ---      &  105.0    \\
\bottomrule
\end{tabular}
\label{tab:optimal_eps_s}
\end{table}

\begin{table}[h!]
\centering
\caption{Optimal $\boldsymbol{\epsilon}$ thresholds for hidden state reversal with hidden dimension permutation, over various Gemma-2-2B-IT and Llama-3.1-8B-Instruct layers.}
\begin{tabular}{ccc}
\toprule
\textbf{Layer} & \textbf{Gemma} & \textbf{Llama} \\
\midrule
1  &    12.5  &   0.5     \\
6  &    25.0    &   3.5     \\
11 &    45.0    &   3.7     \\
16 &    73.0    &   5.2     \\
21 &    118.0   &   6.3     \\
26 &    61.0    &   9.8      \\
32 &    ---      &  30.0    \\
\bottomrule
\end{tabular}
\label{tab:optimal_eps_d}
\end{table}

\begin{table}[h!]
\centering
\caption{Optimal $\boldsymbol{\epsilon}$ thresholds for hidden state reversal with factorized-2D permutation, over various Gemma-2-2B-IT and Llama-3.1-8B-Instruct layers.}
\begin{tabular}{ccc}
\toprule
\textbf{Layer} & \textbf{Gemma} & \textbf{Llama} \\
\midrule
1  &    21.0    &   0.3     \\
6  &    26.0    &   3.0     \\
11 &    47.0    &   9.0      \\
16 &    69.0    &   9.0     \\
21 &    118.0    &  14.0      \\
26 &    51.0    &   14.0      \\
32 &    ---      &  45.0    \\
\bottomrule
\end{tabular}
\label{tab:optimal_eps_sd}
\end{table}
\FloatBarrier

\clearpage

\section{Permuted Setting Attack Algorithms}
\label{appendix:perm_algos}

\begin{algorithm}[h!]
\caption{Attack on Sequence Dimension Permuted LLM Hidden States}
\label{alg:vocab_matching_attack_seq}
\begin{algorithmic}[1]
\INPUT Model $M$, permuted layer $l$ hidden states $\boldsymbol{h} = [h_{\sigma(1)}, h_{\sigma(2)}, ..., h_{\sigma(N)}]$, vocabulary $\mathcal{V}$, proposal model $P$, L1-threshold $\boldsymbol{\epsilon}$%, KV-cache $C$
\OUTPUT Decoded token sequence $\widehat{\boldsymbol{x}} = [\widehat{x}_1, \widehat{x}_2, \ldots, \widehat{x}_N]$
\STATE Initialize empty sequence $\boldsymbol{\widehat{x}} \gets []$
\STATE Initialize set of remaining hidden states $\mathcal{H} \gets \{h_{\sigma(1)}, h_{\sigma(2)}, ..., h_{\sigma(N)}\}$
%\STATE Initialize empty KV-cache $C \gets \{\}$
\FOR{$i = 1$ to $N$}
    \STATE $\mathcal{V}_{\text{ordered}} \gets \text{argsort}(P([\widehat{\boldsymbol{x}}, v] | \widehat{\boldsymbol{x}}))$ \COMMENT{Get ordered vocabulary from proposal model}
    \STATE $\text{min\_dist} \gets \infty$
    \STATE $\text{best\_match} \gets \text{None}$
    \FOR{$v \in \mathcal{V}_{\text{ordered}}$}       
        \STATE $g \gets M_{\leq l}([\widehat{\boldsymbol{x}}, v])$ \COMMENT{Forward pass up to layer $l$} %with KV-caching}
        \FOR{$h \in \mathcal{H}$}
            \STATE $\text{dist} \gets \| g - h|\|_1$ \COMMENT{Calculate L1 distance}
            \IF{$\text{dist} < \text{min\_dist}$}
                \STATE $\text{min\_dist} \gets \text{dist}$
                \STATE $\text{best\_match} \gets v$
                \STATE $\text{best\_h} \gets h$
            \ENDIF
            \IF{$\text{dist} < \boldsymbol{\epsilon}$} 
                \STATE $\widehat{x}_i \gets v$
                \STATE Remove $h$ from $\mathcal{H}$
                %\STATE $C \gets C'$ \COMMENT{Update KV-cache}
                \STATE break
            \ENDIF
        \ENDFOR
    \ENDFOR
    \IF{$\text{min\_dist} \geq \boldsymbol{\epsilon}$}
        \STATE $\widehat{x}_i \gets \text{best\_match}$
        \STATE Remove \text{best\_h} from $\mathcal{H}$
        %\STATE $C \gets C'$ \COMMENT{Update KV-cache}
    \ENDIF
\ENDFOR
\STATE \textbf{return} $\widehat{\boldsymbol{x}}$
\end{algorithmic}
\end{algorithm}

\begin{algorithm}[h!]
\caption{Attack on Hidden Dimension Permuted LLM Hidden States}
\label{alg:vocab_matching_attack_hidden}
\begin{algorithmic}[1]
\INPUT Model $M$, layer $l$ permuted hidden states $\boldsymbol{h} = \left[\pi_1(h_1), \pi_2(h_2), ..., \pi_N(h_N)\right]$, vocabulary $\mathcal{V}$, proposal model $P$, L1-threshold $\boldsymbol{\epsilon}$%, KV-cache $C$
\OUTPUT Decoded token sequence $\widehat{\boldsymbol{x}} = [\widehat{x}_1, \widehat{x}_2, \ldots, \widehat{x}_N]$
\STATE Initialize empty sequence $\boldsymbol{\widehat{x}} \gets []$
%\STATE Initialize empty KV-cache $C \gets \{\}$
\FOR{$i = 1$ to $N$}
    \STATE $\mathcal{V}_{\text{ordered}} \gets \text{argsort}(P([\widehat{\boldsymbol{x}}, v] | \widehat{\boldsymbol{x}}))$ \COMMENT{Get ordered vocabulary from proposal model}
    \STATE $\text{min\_dist} \gets \infty$
    \STATE $\text{best\_match} \gets \text{None}$
    \FOR{$v \in \mathcal{V}_{\text{ordered}}$}
        \STATE $g \gets M_{\leq l}([\widehat{\boldsymbol{x}}, v])$ \COMMENT{Forward pass up to layer $l$} %with KV-caching}
        \STATE $\text{dist} \gets \| \text{sort}(g) - \text{sort}(\pi_i(h_i))|\|_1$ \COMMENT{Calculate L1 distance of sorted vectors}
        \IF{$\text{dist} < \text{min\_dist}$}
            \STATE $\text{min\_dist} \gets \text{dist}$
            \STATE $\text{best\_match} \gets v$
        \ENDIF
        \IF{$\text{dist} < \boldsymbol{\epsilon}$} 
            \STATE $\widehat{x}_i \gets v$
            %\STATE $C \gets C'$ \COMMENT{Update KV-cache}
            \STATE break
        \ENDIF
    \ENDFOR
    \IF{$\text{min\_dist} \geq \boldsymbol{\epsilon}$}
        \STATE $\widehat{x}_i \gets \text{best\_match}$
        %\STATE $C \gets C'$ \COMMENT{Update KV-cache}
    \ENDIF
\ENDFOR
\STATE \textbf{return} $\widehat{\boldsymbol{x}}$
\end{algorithmic}
\end{algorithm}

\begin{algorithm}[h!]
\caption{Attack on Factorized 2D Permuted LLM Hidden States}
\label{alg:vocab_matching_attack_2D}
\begin{algorithmic}[1]
\INPUT Model $M$, permuted layer $l$ hidden states $\boldsymbol{h} = [\pi_1(h_{\sigma(1)}), \pi_2(h_{\sigma(2)}), ..., \pi_N(h_{\sigma(N)})]$, vocabulary $\mathcal{V}$, proposal model $P$, L1-threshold $\boldsymbol{\epsilon}$%, KV-cache $C$
\OUTPUT Decoded token sequence $\widehat{\boldsymbol{x}} = [\widehat{x}_1, \widehat{x}_2, \ldots, \widehat{x}_N]$
\STATE Initialize empty sequence $\boldsymbol{\widehat{x}} \gets []$
\STATE Initialize set of remaining hidden states $\mathcal{H} \gets \{\pi_1(h_{\sigma(1)}), \pi_2(h_{\sigma(2)}), ..., \pi_N(h_{\sigma(N)})\}$
%\STATE Initialize empty KV-cache $C \gets \{\}$
\FOR{$i = 1$ to $N$}
    \STATE $\mathcal{V}_{\text{ordered}} \gets \text{argsort}(P([\widehat{\boldsymbol{x}}, v] | \widehat{\boldsymbol{x}}))$ \COMMENT{Get ordered vocabulary from proposal model}
    \STATE $\text{min\_dist} \gets \infty$
    \STATE $\text{best\_match} \gets \text{None}$
    \FOR{$v \in \mathcal{V}_{\text{ordered}}$}       
        \STATE $g \gets M_{\leq l}([\widehat{\boldsymbol{x}}, v])$ \COMMENT{Forward pass up to layer $l$} %with KV-caching}
        \FOR{$h \in \mathcal{H}$}
            \STATE $\text{dist} \gets \| \text{sort}(g) - \text{sort}(h)|\|_1$ \COMMENT{Calculate L1 distance of sorted vectors}
            \IF{$\text{dist} < \text{min\_dist}$}
                \STATE $\text{min\_dist} \gets \text{dist}$
                \STATE $\text{best\_match} \gets v$
                \STATE $\text{best\_h} \gets h$
            \ENDIF
            \IF{$\text{dist} < \boldsymbol{\epsilon}$} 
                \STATE $\widehat{x}_i \gets v$
                \STATE Remove $h$ from $\mathcal{H}$
                %\STATE $C \gets C'$ \COMMENT{Update KV-cache}
                \STATE break
            \ENDIF
        \ENDFOR
    \ENDFOR
    \IF{$\text{min\_dist} \geq \boldsymbol{\epsilon}$}
        \STATE $\widehat{x}_i \gets \text{best\_match}$
        \STATE Remove \text{best\_h} from $\mathcal{H}$
        %\STATE $C \gets C'$ \COMMENT{Update KV-cache}
    \ENDIF
\ENDFOR
\STATE \textbf{return} $\widehat{\boldsymbol{x}}$
\end{algorithmic}
\end{algorithm}

\clearpage

\section{Expanded Details on Conducting the Attack with a Private Embedding Layer}
\label{appendix:private_embedding}

Here we provide further details on the explicit methods by which an adversary may carry out our attack despite the embedding layer remaining private in the schemes of \citet{yuan2024securetransformerinferenceprotocol}, \citet{luo2024centaurbridgingimpossibletrinity}.

\paragraph{Tied Embedding Case} First, as we described earlier, for many models -- such as the Gemma family -- the embedding matrix is simply the transpose of the language-modeling head, whose permutation (in row and column dimensions) is known to $P_2$. Therefore, the vocabulary embedding vectors to search over in this case are simply permuted columns of this permuted language-modeling head -- and these permutations can be uncovered by matching against the permuted input embedding vectors received from the user at inference. 

Explicitly, denoting $W$ as the original $\mathbb{R}^{V \times d}$ embedding matrix, $P_2$ has access to the permuted language-modeling head $\pi_d W^T \pi_V \in \mathbb{R}^{d \times V}$, where $\pi_V,\pi_d$ are $V \times V,d\times d$ permutation matrices. In inference, the user $P_3$ first applies a $d \times d$ permutation $\pi$ on the input embeddings $e_1,\ldots,e_N \in \mathbb{R}^d$; these are rows of $W$. Therefore, $P_2$ sees permuted embedding vectors $e_1 \pi, \ldots, e_N\pi \in \mathbb{R}^{d}$. Now, assuming the uniqueness of \textit{sorted} rows of $W$, each $e_i \pi$ can be obtained by applying the permutation $\pi \pi_d^{-1}$ on exactly one column of $\pi_d W^T \pi_V$. Thus $P_2$ can recover $\pi \pi_d^{-1}$ by looking for a sorted match between the columns of $\pi_d W^T \pi_V$ and each $e_i \pi$. Once obtained, they can compute $\pi \pi_d^{-1} \pi_d W^T \pi_V = \pi W^T \pi_V$, whose columns are precisely all $\pi$-permuted vocabulary embeddings. With these, because the altered transformer forward pass is carried out on $\pi$-permuted embeddings, $P_2$ can carry out our attack on any permuted layer $l$ hidden states it obtains. 

To confirm the plausibility of the above, we examined the embedding matrices of Gemma, Llama and Mistral models and found that it is indeed the case that for these modern LLM families, the rows of $W$ are unique even when sorted.

\paragraph{Non-Tied Case} Even if the language-modeling head is not the transpose of the embedding matrix, $P_2$ can collect the set of sorted input embeddings over the course of many inference requests. After sufficiently many calls, they can then perform our attack by iterating through this collection of embeddings, permuting them to match the initial permuted input embeddings. The only difference in this case is that $P_2$ must wait for more inference requests in order to carry out the attack, rather than being able to perform it immediately.

The final step to decoding by the adversary is then mapping the embeddings back into tokens. If the tokenizer is not publicly revealed, this may seem difficult at first -- but note that this essentially constitutes a simple substitution cipher. Again, by collecting data over many queries and using simple methods such as frequency analysis and positional information, $P_2$ can learn to decode this into the original tokens; substitution ciphers are in general easily broken given sufficient data.

\section{Proof of \cref{theorem:1dproj}}
\label{appendix:proof_1dproj}

Here, we provide a proof of our \cref{theorem:1dproj} given in \cref{sec:distcorr_theory}. At a high level, we show that the inequality below is true whenever a randomly weighted sum of a vector far exceeds its L2 norm, and this holds whenever a weight coordinate is sufficiently large.

    \begin{theoremrestated}
    \label{theorem:1dproj_appendix}
        Let $k>0$. Suppose random weights $\bm{w} \in \mathbb{R}^d$ are drawn from a $d$-variate spherically symmetric distribution $\mathcal{D}$. Then any $\bm{x},\bm{y} \in \mathbb{R}^d$, we have the absolute difference of $\bm{w}$-weighted sums of $\bm{x}$ and $\bm{y}$ exceeds the L1 distance between $\bm{x}$ and $\bm{y}$ by a factor $\geq k$, meaning
        \begin{equation}\label{eqn:random_l1}\left|\sum_{i=1}^d w_i x_i - \sum_{i=1}^d w_i y_i\right| \geq k \sum_{i=1}^d |x_i-y_i|,\end{equation}
        with probability $\geq P_{\bm{\gamma} \sim \mathcal{D}} (|\gamma_1| \geq k\sqrt{d})$.
    \end{theoremrestated}
    
    \begin{proof}
        Denote $\bm{z} = \bm{x}-\bm{y}$. Observe that
        $$\left|\sum_{i=1}^d w_i x_i - \sum_{i=1}^d w_i y_i\right| = \left|\sum_{i=1}^d w_i (x_i-y_i)\right| = \left|\sum_{i=1}^d w_i z_i \right| = |\bm{w}^T \bm{z}|.$$
        Thus, \cref{eqn:random_l1} is equivalent to $|\bm{w}^T\bm{z}| \geq k\|\bm{z}\|_1$. Then, from the standard bound $\|\bm{z}\|_1 \leq \sqrt{d} \|\bm{z}\|_2$, which can be proven by an application of Cauchy-Schwarz, we see that \cref{eqn:random_l1} holds whenever 
        \begin{equation}\label{eqn:random_l1_red}|\bm{w}^T\bm{z}| \geq k\sqrt{d}\|\bm{z}\|_2.\end{equation}
        We now aim to compute the probability of the above event. Choose a $d\times d$ orthogonal matrix $Q$ such that $\bm{z}_q:=Q\bm{z} \in \mathbb{R}^d$ only has a nonzero coordinate $L$ in its first position, i.e. $\bm{z}_q=(L,0,\ldots,0)$. By orthogonality and the fact that $\mathcal{D}$ is spherically symmetric, we see $\bm{w}_q:=Q\bm{w}$ has distribution $\mathcal{D}$.  Furthermore, orthogonal linear transformations are length-preserving (by L2 norm), so we have $\|\bm{z}_q\|_2=\|Q\bm{z}\|_2 = \|\bm{z}\|_2=|L|$. In fact, as $Q^TQ=I$, observe that $\bm{w}^T \bm{z} = \bm{w}^T Q^T Q \bm{z} = (Q\bm{w})^T (Q\bm{z}) = \bm{w}_q^T \bm{z}_q$. Hence, \cref{eqn:random_l1_red} becomes
        $$|\bm{w}_q^T \bm{z}_q|  = |L| |(\bm{w}_q)_1| \geq k|L|\sqrt{d}.$$
        This is equivalent to saying the first coordinate of $\bm{w}_q$ has magnitude at least $k\sqrt{d}$. But we showed $\bm{w}_q$ has distribution $\mathcal{D}$, so the probability that \cref{eqn:random_l1_red} holds is precisely $P_{\bm{\gamma} \sim \mathcal{D}} (|\gamma_1|\geq k\sqrt{d})$. This is therefore a lower bound on the probability that \cref{eqn:random_l1} holds, since we showed \cref{eqn:random_l1} holds whenever \cref{eqn:random_l1_red} does.
    \end{proof}

\section{Proof of
\cref{theorem:discorr}}
\label{appendix:proof_discorr}

We now provide a proof of our \cref{theorem:discorr} from \cref{sec:distcorr_theory}. The idea is to construct a $\delta$-reconstructible pair with low distance correlation by using absolute values of symmetric variables, and then form a non-$\delta$-reconstructible pair with high distance correlation by using highly correlated normal variables.

\begin{theoremrestated}
    For any $\delta>0$, there exist random variables $W,X,Y,Z$ such that $\text{Discorr}(W,X) > \text{Discorr}(Y,Z)$, the pair $(W,X)$ is not $\delta$-reconstructible, and the pair $(Y,Z)$ is $\delta$-reconstructible. 
    \label{theorem:discorr_appendix}
\end{theoremrestated}

\begin{proof}
    Define independent random variables $W,\varepsilon \sim \mathcal{N}(0,1)$. Let $Y$ come from an arbitrary symmetric distribution about zero with support $[-\delta,\delta]$, and construct
    $$X = \rho W + \sqrt{1-\rho^2} \varepsilon, \quad Z = |Y|$$
    where $1>\rho > 0.945$. Using standard properties of normal random variables, one can see $X \sim \mathcal{N}(0,1)$, and the correlation between $X$ and $W$ is $\rho$. Thus, by Theorem 7 in \citet{szekely2007measuring}, which lower bounds distance correlation of standard normals in terms of (Pearson) correlation, we have $\text{DisCorr}(W,X) > 0.89\rho > 0.841$. Furthermore, by Theorem 1 in \citet{edelmann2021relationships}, which upper bounds the distance correlation of a symmetric random variable and its absolute value, we have $\text{DisCorr}(Y,Z) \leq 2^{-1/4} < 0.841$. Therefore, we have $\text{DisCorr}(W,X) > \text{DisCorr}(Y,Z)$.
    
    Now, we claim that $(W,X)$ is not $\delta$-reconstructible. To see this, note $(W,\epsilon) \sim \mathcal{N}(0,I)$ by independence, so the linear transformation $(W,\epsilon) \mapsto (W,X)$ can be seen to induce the joint distribution
    $$(W,X) \sim \mathcal{N}\left(0,\Sigma=\begin{pmatrix}1 & \rho \\ \rho & 1\end{pmatrix}\right).$$
    From the standard conditional Gaussian formula, one obtains $W|(X=x) \sim \mathcal{N}(\rho x, \sqrt{1-\rho^2})$. Thus, for any estimator $f(X)$ of $Y$, we have for each $x$ that
    \begin{align*}
        P\left(|W-f(X)| \leq \delta | X=x\right) &\leq P\left(|W-\rho x| \leq \delta | X=x\right) = 2\Phi\left(\frac{\delta}{\sqrt{1-\rho^2}}\right) -1= c < 1
    \end{align*}
    where $c$ is a constant dependent on $\rho,\delta$, and $\Phi$ is the normal CDF. Here, the first inequality holds as $W|(X=x)$ is a normal distribution: this means $P\left(|W-f(X)| \leq \delta | X=x\right)$, the integral of the corresponding normal PDF over $(f(x)-\delta,f(x)+\delta)$, is upper bounded\footnote{This directly follows from the fact that an integral of a zero-centered normal (or generally any unimodal symmetric distribution) over a fixed-size interval is maximal when that interval is zero-centered. This is a standard fact: see the first sentence in \citep{anderson1955integral}, for example.} by its integral over the same-size mean-centered interval $(\rho x-\delta,\rho x + \delta)$, which is precisely $P\left(|W-\rho x| \leq \delta | X=x\right)$. Finally, taking the expectation of the above bound over $X$ and applying the law of total expectation, we get $P(|W-f(X)| \leq \delta) \leq c < 1$. Since $f(X)$ was chosen arbitrarily, this shows $(W,X)$ is not $\delta$-reconstructible\footnote{In fact, it shows something stronger: the optimal estimator's probability of reconstructing $W$ up to an absolute error of $\delta$ is upper bounded by $c$. As $\delta \to 0$, the value of $c$ actually approaches $2\Phi(0)-1=0$.}, as required for the claim.
    
    However, $(Y,Z)$ is certainly $\delta$-reconstructible. Because $|Y| \leq \delta$ almost always, we see $f(Z)=0$ always estimates $Y$ within a $\delta$-threshold. Hence, we have our desired counterexample. 
\end{proof}

\section{Scalability of Attack}
\label{appendix:scalability}

To assess the scalability of our attack with respect to model size, we conducted additional experiments across a range of model scales, from 1 billion to 27 billion parameters. Table~\ref{tab:scaling_experiment} summarizes the results.

\begin{table}[h!]
\centering
\caption{Average attack time (in seconds) over 10 decodings for various model sizes.}

\label{tab:scaling_experiment} 
\resizebox{\textwidth}{!}{
\begin{tabular}{cccc}
\toprule
Model Name & Model Size (Parameters) & Vocabulary Size & Average Attack Time (s) \\
\midrule
Llama-3.2-1B-Instruct & 1B & 128,256 & 49 \\
Gemma-2-2B-IT & 2B & 256,000 & 124 \\
Llama-3.1-8B-Instruct & 8B & 128,256 & 69 \\
Gemma-2-27B-IT ($\epsilon=30$) & 27B & 256,000 & 304 \\
Gemma-2-27B-IT ($\epsilon=40$) & 27B & 256,000 & 124 \\
\bottomrule
\end{tabular}
}
\end{table}

The attack time remains practical across all evaluated model sizes, typically on the order of minutes for perfect decoding of length 100 prompts. We observe that the computational cost is primarily a function of the vocabulary size and the choice of $\epsilon$, rather than the total number of model parameters. Specifically, models with larger vocabularies (e.g., 256,000 tokens) exhibit proportionally longer attack times compared to models with smaller vocabularies (e.g., 128,256 tokens), regardless of parameter count. While a poorly chosen $\epsilon$ leads to longer runtimes, it does not fundamentally impede the attack. These results demonstrate that the attack scales favorably to larger models, including recent LLMs with tens of billions of parameters.

\section{Noising method performance}
\label{appendix:noising_method}

Below, we provide exact (not only the maximum) ROUGE scores across layers $1,6,11,16,21,26$, for all methods of noising discussed in \cref{sec:noised_decoding}. \cref{tab:gaussian001_U}, \cref{tab:gaussian001_S}, \cref{tab:gaussian001_D} show these results. We also provide a complete breakdown of LiveBench scores per category in \cref{tab:gemma_noised_livebench}.

\begin{table}[h!]
\centering
\caption{The ROUGE scores of decoded texts with added noise and no permutation.}
\begin{tabular}{cccccc}
\toprule
\textbf{Layer}  & \textbf{$\sigma=10^{-2}$} & \textbf{$\sigma=10^{-1}$} & Random Emb & 8-bit quantization & 4-bit quantization\\
\midrule
1  &    0.9263  & 0.9177 & 0.9309 & 0.8901 & 0.8844\\
6  &    0.9273  & 0.3271 & 0.9340 & 0.8726 & 0.8652\\
11 &    0.9070  & 0.0856 & 0.8170 & 0.8943 & 0.8764\\
16 &    0.9175  & 0.0587 & 0.7552 & 0.8620 & 0.8669\\
21 &    0.9232  & 0.0977 & 0.8247 & 0.8834 & 0.8839\\
26 &    0.9070  & 0.0485 & 0.6257 & 0.8751 & 0.8771\\
\bottomrule
\end{tabular}
\label{tab:gaussian001_U}
\end{table}

\begin{table}[h!]
\centering
\caption{The ROUGE scores of decoded texts with added noise and sequence dimension permutation.}
\begin{tabular}{cccccc}
\toprule
\textbf{Layer}  & \textbf{$\sigma=10^{-2}$} & \textbf{$\sigma=10^{-1}$} & Random Emb & 8-bit quantization & 4-bit quantization\\
\midrule
1  &    0.0696  & 0.0000 & 0.1683 & 0.8167 & 0.8157\\
6  &    0.0354  & 0.0000 & 0.0418 & 0.8236 & 0.8409\\
11 &    0.0278  & 0.0011 & 0.0337 & 0.8479 & 0.8138\\
16 &    0.0133  & 0.0023 & 0.0202 & 0.8568 & 0.8116\\
21 &    0.0136  & 0.0051 & 0.0321 & 0.8283 & 0.8250\\
26 &    0.0096  & 0.0096 & 0.0236 & 0.8236 & 0.7956\\
\bottomrule
\end{tabular}
\label{tab:gaussian001_S}
\end{table}

\begin{table}[h!]
\centering
\caption{The ROUGE scores of decoded texts with added noise and hidden dimension permutation.}
\begin{tabular}{cccccc}
\toprule
\textbf{Layer}  & \textbf{$\sigma=10^{-2}$} & \textbf{$\sigma=10^{-1}$} & Random Emb & 8-bit quantization & 4-bit quantization\\
\midrule
1  &    0.0669  & 0.0000 & 0.1945 & 0.7544 & 0.7497\\
6  &    0.0353  & 0.0000 & 0.0359 & 0.6696 & 0.6420\\
11 &    0.0301  & 0.0009 & 0.0300 & 0.6667 & 0.8138\\
16 &    0.0166  & 0.0018 & 0.0144 & 0.6325 & 0.8116\\
21 &    0.0164  & 0.0036 & 0.0153 & 0.5029 & 0.8250\\
26 &    0.0116  & 0.0101 & 0.0114 & 0.3848 &0.7956\\
\bottomrule
\end{tabular}
\label{tab:gaussian001_D}
\end{table}

\begin{table}[h!]
\centering
\caption{The ROUGE scores of decoded texts with added noise and factorized 2D permutation.}
\begin{tabular}{cccccc}
\toprule
\textbf{Layer}  & \textbf{$\sigma=10^{-2}$} & \textbf{$\sigma=10^{-1}$} & Random Emb & 8-bit quantization & 4-bit quantization\\
\midrule
1  &    0.0675  & 0.0000 & 0.1919 & 0.7328 & 0.7146\\
6  &    0.0346  & 0.0000 & 0.0361 & 0.6075 & 0.5820\\
11 &    0.0273  & 0.0016 & 0.0297 & 0.4916 & 0.5753\\
16 &    0.0182  & 0.0027 & 0.0140 & 0.3731 & 0.5701\\
21 &    0.0196  & 0.0044 & 0.0151 & 0.3845 & 0.5568\\
26 &    0.0116  & 0.0120 & 0.0117 & 0.3496 & 0.5564\\
\bottomrule
\end{tabular}
\label{tab:gaussian001_SD}
\end{table}

\begin{table}[h!]
\centering
\small
\setlength{\tabcolsep}{4pt} % Reduce column spacing
\caption{Performance of Gemma-2-2B-IT on LiveBench with added noise.}
\begin{tabular}{cccccccc}
\toprule
\textbf{Method} & \textbf{Avg.} & \textbf{Coding} & \multicolumn{1}{c}{\textbf{Data}} & \multicolumn{1}{c}{\textbf{Instruction}} & \textbf{Language} & \textbf{Math} & \textbf{Reasoning} \\ 
& & & \multicolumn{1}{c}{\textbf{Analysis}} & \multicolumn{1}{c}{\textbf{Following}} & & & \\
\midrule
Baseline (no noise) & 20.7 & 9.4 & 26.1 & 48.9 & 15.2 & 13.1 & 11.3\\
Gaussian, \textbf{$\sigma=10^{-2}$} & 21.0 & 11.1 & 27.4 & 51.2 & 13.7 & 13.4 & 9.3\\
Gaussian, \textbf{$\sigma=10^{-1}$}& 1.2 & 0.0 & 0.0 & 6.9 & 0.4 & 0.0 & 0.0\\
Random emb. prefix & 21.3 & 8.8 & 27.5  & 50.1 & 16.1 & 13.6 & 12.0\\
8-bit quantization & 20.2 & 8.8 & 27.1 & 49.2 & 13.3 & 13.0 & 10.0 \\
4-bit quantization & 19.1 & 6.5 & 25.5 & 50.5 & 9.5 & 10.9 & 12.0 \\
\bottomrule
\end{tabular}
\label{tab:gemma_noised_livebench}
\end{table}

\end{document}